\documentclass[]{article}
\usepackage[a4paper]{geometry}
\usepackage[utf8]{inputenc}
\usepackage[colorlinks=true, allcolors=blue, citecolor=gray]{hyperref}
\usepackage{amsmath,amsfonts,amssymb}
\usepackage{float, graphicx, cleveref, setspace, relsize}
\usepackage[inkscapelatex=false]{svg} 
\usepackage[numbers]{natbib} 
\usepackage{bm} 
\usepackage[utf8]{inputenc}
\usepackage{pdfpages} 
\DeclareUnicodeCharacter{2212}{-} 
\usepackage{authblk} 
\usepackage{blindtext}
\usepackage{subfiles} 
\usepackage{soul} 
\usepackage{chngcntr} 

\title{\textbf{Fewer Fractions, Equieffective Plans: An optimization framework based on equieffective dose}}

\author[a,b,c,*]{Suryakant Kaushik}
\author[a]{Jakob Ödén}
\author[a]{Albin Fredriksson}
\author[b,c]{Iuliana Toma-Dasu}

\affil[a]{RaySearch Laboratories AB (Publ), Stockholm, Sweden}
\affil[b]{Department of Physics, Medical Radiation Physics, Stockholm University, Stockholm, Sweden}
\affil[c]{Department of Oncology and Pathology, Medical Radiation Physics, Karolinska Institutet, Stockholm, Sweden}
\affil[*]{Corresponding author: Suryakant Kaushik at RaySearch Laboratories AB (Publ), Stockholm, Sweden, E-mail address: suryakant.kaushik@raysearchlabs.com}

\date{}

\begin{document}

\maketitle


\noindent \textbf{Manuscript word count:} 3531 \\
\textbf{Abstract word count:} 500 \\

\section*{\hfil Abstract \hfil}
    \textbf{Background}: Hypofractionation shortens the overall treatment course by reducing the number of fractions and offers the advantage of potentially increasing patient access to radiotherapy, an aspect that is particularly relevant for proton therapy given its limited availability. However, employing non-standard fractionation requires particular caution before clinical implementation, and treatment planners often need to manually modify the prescription and dose constraints to account for non-linear biological effects. Such workflows can be time-consuming and may introduce additional complexity or risk for error.
        
    \noindent \textbf{Purpose}: To develop an extension within a treatment planning system that enables the generation of hypofractionated plans based on equieffective dose, $EQD_{n_{ref}}$, where $n_{ref}$ denotes the fixed reference number of fractions. Our method allows for direct optimization of $EQD_{n_{ref}}$ for an alternative fractionation deliverable in $n$ fractions, such that it is equieffective with the reference fractionation for a specified biological endpoint. This also eliminates the need for manual recalculation of prescriptions and constraints when changing fractionation.
    
    \noindent \textbf{Methods}: A research build of RayStation v2025 was extended to optimize plans based on $EQD_{n_{ref}}$. Anonymized data from three head and neck patients were included in this study. Primary and secondary clinical target volumes ($\text{CTV}_p$/$\text{CTV}_n$) were prescribed 70 Gy and 54 Gy, respectively, in 35 fractions. The $\text{CTV}_p$ clinical goals were to achieve an average dose and a D98 greater than 100\% and 98\% of the prescribed dose, respectively. The targets, spinal cord, and brainstem were robustly optimized assuming an isotropic setup uncertainty of $\pm3.0$ mm and a density uncertainty of $\pm3.5\%$. Equieffective hypofractionated plans were generated for each patient using 34, 32, 30 and 25 fractions, while all other plan settings were kept unchanged from those of the nominal plan. A sensitivity analysis was also performed to assess the impact of $\alpha/\beta$ uncertainty on the resulting dose distribution.   
    
    \noindent \textbf{Results}: All equieffective hypofractionated plans maintained target coverage and satisfied the mandatory clinical goals for organs-at-risk (OARs). Across all patients, the mean dose to $\text{CTV}_p$ was 70.5 Gy (SD 0.08 Gy) when all plans were reported in $EQD_{n_{ref}}$. In addition, all scenarios satisfied the robustness criterion of D$95 \geq 95\%$ of the prescribed dose for each patient. In the direct one-to-one comparison between the standard and hypofractionated scenarios, the DVH parameters indicated an average $\Delta EQD_{n_{ref}}$ of less than 0.2 Gy for $\text{CTV}_p$, with no difference observed for $\text{CTV}_n$. For parallel OARs, the median dose difference between the hypofractionated plan (down to 30 fractions) and the standard fractionation plan was mostly less than 1\% of the prescribed dose. Uncertainty in the $\alpha/\beta$ values led to the largest variation in the 25-fraction hypofractionated regimen, with absolute $\Delta EQD_{n_{ref}}$ values up to 0.88 Gy for the targets and as high as 0.63 Gy for the spinal cord and brainstem. 

    \noindent \textbf{Conclusions}: An $EQD_{n_{ref}}$-based framework enables the generation of equieffective hypofractionated treatment plans with minimal difference from conventionally fractionated plans. Using this approach, it is feasible to reduce the treatment fractions while ensuring target coverage and compliance with mandatory OAR clinical goals, without substantially deviating from existing clinical protocols.

\section*{\hfil Keywords \hfil}
    hypofractionation, equivalent dose, radiobiological optimization \\


\newpage
\section{Introduction}
    In radiotherapy, hypofractionation refers to a regimen in which fewer fractions are delivered, requiring a higher dose per fraction than in conventional fractionation to maintain an equieffective treatment \citep{santos2022role}. By reducing the number of fractions, this strategy shortens the total treatment course with the potential to reduce the cost of treatment, improve patient adherence and comfort by decreasing travel time, and may help limit tumor repopulation time \citep{santos2022role, bocha2025hypofractionated, wilkes2006accessing}. Advances in both the hardware and software aspects of external beam radiotherapy have made it possible to deliver conformal radiation with a high dose per fraction. This, in turn, has contributed to the growing use of hypofractionation across multiple anatomical sites. For certain indications, particularly early\textendash stage breast and prostate cancer, hypofractionated schedules have become standard of care, with increasing adoption also in patients requiring regional nodal irradiation \citep{haviland2013uk, catton2017randomized}. In addition, hypofractionation offers the advantage of potentially increasing patient access to radiotherapy, an aspect that is particularly relevant for proton therapy given its limited availability and constrained treatment capacity.
    
    Hypofractionation has broadly been categorized into ultra-hypofractionation and moderate-hypofractionation. Ultra-hypofractionation delivers a treatment in 5 fractions or less, with a very high dose per fraction, usually exceeding 5 Gy per fraction. This approach is commonly used in prostate cancer \citep{nilsson2026ultra}, breast cancer \citep{chakraborty2024ultra} and forms the basis for stereotactic techniques such as stereotactic body radiation therapy and fractionated stereotactic radiosurgery, which represent highly precise implementations of ultra-hypofractionated radiotherapy \citep{kuncman2025definition}. Moderate-hypofractionation substantially reduces the number of fractions compared to conventional fractionation and delivers a treatment in 3\textendash 4 weeks with 2.5\textendash 3.5 Gy per fraction, which is already established as a standard of care for several treatment sites \citep{krug2021moderate, dearnaley2025moderate}. However, its application at other sites\textemdash such as in the head and neck (HN)\textemdash remains limited, largely due to the lack of high quality evidence and ongoing concerns about potential long term toxicities as well as the risk of increased acute mucosal reactions \citep{eom2024hypofractionated, moon2025hypofractionated}. For moderate- and ultra-hypofractionated regimens, numerous guidelines and recommendations are now available to support the treatment planning \citep{emami1991tolerance, marks2010guest, bisello2022dose, grimm2021high}. However, when employing fraction sizes that differ substantially from those of studies in the published literature or non-standard fractionation, particular caution is required before clinical implementation \citep{marks2010use}.

    However, relatively little attention has been paid to mild-hypofractionation\textemdash typically 2.1\textendash 2.5 Gy per fraction\textemdash as a means of making fine-grained adjustments to the total number of fractions (a reduction of approximately 1–5 fractions) relative to conventional fractionation, and the regimen has remained less well characterized. Such small adjustments in fractionation may offer practical advantages, for example, in situations involving missed fractions, machine downtime, public holidays, travel-related burdens for patients or caregivers, or more broadly in the context of time-intensive adaptive treatment workflows. A practical strategy to manage unplanned interruptions in radiation is to increase the dose per fraction, which is well known to compensate for missed treatment days \citep{bese2007effects}. A recent example is the increased interest in accelerated treatment schedules for HN cancer that emerged during the COVID-19 pandemic \citep{vreugdenhil2021hypofractionated}. Data from HN cohorts, including Princess Margaret's experience, as well as global analyses of radiotherapy practice, showed that moderately hypofractionated regimens, with fraction sizes up to  2.4 Gy, were adopted as a necessary clinical strategy rather than as a purely radiobiological preference \citep{huang2020hypofractionated, thomson2020314, portaluri2020hypofractionation}. In this context, mild-hypofractionation can serve as a middle ground with improved logistic efficiency while avoiding the steep toxicity risks associated with more aggressive regimens. However, merely scaling the total dose of a treatment plan fails to account for the nonlinear dependence of biological effect on dose per fraction, as represented by the linear-quadratic (LQ) model \citep{fowler1989linear, mcmahon2019linear}. As a result, treatment planners often need to perform additional calculations to account for non-linear biological effects on target dose and organ-at-risk (OAR) constraints when deviating from conventional fractionation. Such workflows can be time-consuming and may introduce additional complexity or potential for error.
    
    The key contribution of this work is the use of an equieffective dose, $EQD_{n_{ref}}$, as an optimization quantity. In contrast to commonly used $EQD_{X}$ \citep{bentzen2012bioeffect}, which uses a fixed reference fraction size of $X$ Gy, $EQD_{n_{ref}}$ is defined using a fixed reference number of fractions, $n_{ref}$. It denotes the dose delivered in $n_{ref}$ fractions that is equieffective with the test fractionation regimen delivered in $n$ fractions for a specified clinical or biological endpoint. Although equieffective dose formulations have previously been employed for plan evaluation and comparison \citep{niebuhr2021biologically}, the integration directly into the optimization has been limited. Our approach enables direct optimization and control of the equieffective dose across fractionation schemes within the treatment planning process. It also eliminates the need for manual recalculation or iterative adjustment of prescriptions and constraints when modifying fractionation, and provides a consistent radiobiological framework for plan generation, evaluation, and comparison. To our knowledge, this represents the first systematic application of $EQD_{n_{ref}}$-based optimization for exploring and generating equieffective hypofractionated treatment plans relative to a clinically established reference.

    In this study, we developed an extension within a treatment planning system (TPS) that enables the generation and optimization of treatment plans based on $EQD_{n_{ref}}$. Starting from a clinically used optimization problem formulated in dose for a conventional fractionation plan delivered in $n_{ref}$ fractions, the same optimization problem can be reformulated in terms of $EQD_{n_{ref}}$ and applied to different test fractionation schedules, including mild-hypofractionated schedules, by changing only the number of test fractions, $n$. This is possible because, for the conventional schedule, $EQD_{n_{ref}}$ is equal to the physical dose. The approach therefore eliminates the need to explicitly precalculate equieffective dose prescriptions and constraints for each fractionation schedule. The resulting plans were then benchmarked against conventional fractionation plans to assess differences in plan robustness and target/OAR dose metrics. Finally, a sensitivity analysis was conducted to investigate how uncertainty in the LQ model parameter influences the resulting dose distribution.

\section{Material and Methods}
    \subsection{Clinical data and conventional treatment planning}
        The study included anonymized data from three HN cancer patients. For each patient, a planning computed tomography scan was available, in which the radiation oncologist had contoured the targets and organs at risk (OARs). A constant value of proton relative biological effectiveness (RBE) of 1.1 was adopted in this work as recommended for proton therapy \citep{paganetti2019report, oden2025beyond}, and all subsequent references to ``Gy" in this study indicate ``Gy (RBE)". Primary and secondary clinical target volumes ($\text{CTV}_p/\text{CTV}_n$) were prescribed doses of 70 Gy and 54 Gy, respectively, delivered in 35 fractions using a simultaneous integrated boost technique. The clinical goals were to achieve an average dose and a D98 (where D$x$ denotes the minimum dose delivered to $x$\% of the volume) to $\text{CTV}_p$ greater than 100\% and 98\% of the prescribed dose, respectively.  

        For this study, a research build of RayStation v2025 TPS was used. For each patient, a nominal intensity-modulated proton therapy (IMPT) plan was generated. Each plan comprised four beams (left anterior oblique, right anterior oblique, left posterior oblique, and right posterior oblique) and used a 4 cm water-equivalent thickness range shifter for the anterior beams. The air gap between the snout and the patient’s body was kept at 5 cm. Spinal cord, brainstem and targets were robustly optimized assuming an isotropic setup uncertainty of $\pm3.0$ mm and a density uncertainty of $\pm3.5\%$. For $\text{CTV}_p$, the robustness criterion was that D95 (and D98 when achievable) should exceed 95\% of the prescribed dose.
        
        To formally describe the optimization framework, let $f_k$, with $k \in K$, represent the optimization function in the nominal IMPT plan and $K$ is an index set defined for the total number of optimization functions in a plan. These $f_k$ were formulated based on the conventional fractionation regimen and each was assigned a nonnegative weighting factor $w_k$. Let $R \subseteq K$ be the set of indices corresponding to the objectives that are subject to robust optimization. The worst-case scenario approach, also referred to as a minimax approach, was then defined as the robust optimization problem shown in Equation \ref{minimaxproblem} \citep{fredriksson2013robust}.
        
        \begin{align}
            \begin{split}
                \underset{\psi \in \Psi}{\mathrm{minimize}} \quad  \sum_{k \in K \setminus R} w_k f_k(D(\psi)) + \underset{s \in S}{\mathrm{max}} \left(\sum_{k \in R} w_k f_k(D^s(\psi))\right)
            \end{split}
            \label{minimaxproblem}
        \end{align}
        
         where $D(\psi)$ denotes the plan dose in the treatment plan, which depends on the spot weight vector $\psi$ and adheres to the machine’s physical constraints specified by the feasible set $\Psi$, and $D^s(\psi)$ indicates the dose in scenario $s \in S$, corresponding to a given combination of setup and density uncertainty realizations.

         The final dose was computed using the Monte Carlo algorithm with a statistical uncertainty of $0.5\%$. The fractionation pattern of the nominal IMPT plan was taken as the standard fractionation in this study, thereby setting the reference number of fractions, $n_{ref}$, to 35.
        
    \subsection{Equieffective hypofractionated treatment planning}
         A hypofractionated plan refers to a regimen in which the total plan dose, $D$, is delivered over a reduced number of fractions, $n$. The goal is to find an equieffective dose, $EQD_{n_{ref}}$, with $n_{ref}$ fractions, from the current hypofractionated plan dose $D$, such that both doses result in the same probability of achieving the specified biological end point \citep{bentzen2012bioeffect}. This is obtained by setting the survival fractions of the two regimens equal using the non-linear dose–response relationship described by the LQ model, as shown in Equation \ref{survival_equivalence}.
         
        \begin{equation} \label{survival_equivalence}
            -\ln(\text{survival fraction}) = \left( \alpha D + \beta \frac{D^2}{n} \right) = \left( \alpha EQD_{n_{ref}} + \beta \frac{EQD_{n_{ref}}^2}{n_{ref}} \right)            
        \end{equation}
        
        Rearranging the terms in equation \ref{survival_equivalence} yields $EQD_{n_{ref}}$ as expressed in Equation \ref{reference_dose}.
        
        \begin{equation} \label{reference_dose}
            EQD_{n_{ref}}(D, n, \frac{\alpha}{\beta}) = n_{ref} \Bigg[- \frac{1}{2} \bigg( \frac{\alpha}{\beta} \bigg) + \sqrt{\frac{1}{4} \bigg( \frac{\alpha}{\beta} \bigg)^2 + \frac{1}{n_{ref}} \bigg(\frac{\alpha}{\beta} D + \frac{D^2}{n} \bigg)} \Bigg]
        \end{equation}
        where, $\alpha/\beta$ is the LQ model parameter indicating fractionation sensitivity and was set to 10 Gy for tumors and 3 Gy for OARs. In our study, $n_{ref}$ was fixed at 35 fractions and in case $n = n_{ref}$, $EQD_{n_{ref}}$ will reduce to the plan dose $D$.   

        The research version of RayStation was extended to optimize treatment plans based on $EQD_{n_{ref}}$ and allows for region-of-interest (ROI)-specific parameters. Each optimization function $f_k$ associated with a given ROI was evaluated independently using $EQD_{n_{ref}}$ with its ROI-specific parameter set; consequently, overlap among ROIs does not introduce uncertainty. An equieffective hypofractionated plan can be generated by duplicating the previously created nominal IMPT plan and modifying only the number of fractions as desired, while leaving all other aspects—such as plan settings, robustness parameters, and optimization function settings—unchanged. Consequently, each optimization function $f_k$  was evaluated with respect to $EQD_{n_{ref}}$ instead of $D$.

        Equieffective treatment plans were created using 34, 32 and 30 fractions, representing mild hypofractionation, together with a 25-fraction schedule included as a more pronounced hypofractionation scenario to test the framework over a wider fractionation range. No dose scaling was performed following the final dose calculation.

    \subsection{Treatment plan evaluation}
        For evaluation, the plan doses of the hypofractionated plans were converted to $EQD_{n_{ref}}$ ($n_{ref}=35$) using Equation \ref{reference_dose}, allowing comparison between fractionation schemes. Subsequently, planned dose and $EQD_{n_{ref}}$ of the hypofractionated plan were compared with the nominal IMPT plan dose. To avoid geometric overlap and the influence of the steep dose gradient surrounding the primary target when assessing the secondary target, $\text{CTV}_n$ was cropped to maintain a 1 cm separation from $\text{CTV}_p$, and the resulting ROI, termed $\text{CTV}^{crop}_n$, was used for the evaluation.

        Doses were calculated for 27 scenarios per plan. These comprised nine setup conditions, namely the nominal position and the eight combinations of $\pm3$ mm shifts along the three principal axes, each evaluated for the nominal density and density variations of $\pm3.5\%$. Each scenario dose was converted to $EQD_{n_{ref}}$. All robustness evaluations were subsequently carried out using the $EQD_{n_{ref}}$ doses and the corresponding dose-volume histogram (DVH) parameters. For each patient, various dose and DVH parameters obtained from the scenario doses were compared across different fractionation schemes.

    \subsection{\texorpdfstring{$\alpha/\beta$}{alpha/beta} sensitivity analysis}
        The $\alpha/\beta$ value is a key parameter in the LQ model that describes fractionation sensitivity; however, its exact value is uncertain and may vary depending on the assumptions made. Therefore, a sensitivity analysis was carried out by varying the $\alpha/\beta$ values with respect to those adopted for plan optimization. For the targets, an uncertainty interval of $\pm25\%$ was assumed \citep{bolt2021quantification}, whereas for the OARs, additional $\alpha/\beta$ values of 2, 4, and 5 were considered \citep{quashie2023obtaining, book2009joiner, stuschke1999fractionation}. For each hypofractionated treatment plan, the planned dose was converted to $EQD_{n_{ref}}(D)$ using these $\alpha/\beta$ values, and the resulting doses were evaluated and compared using several DVH based parameters.

\section{Results}
    \subsection{Comparison with nominal plan}
\begin{figure}[H]
          \centering
          \includegraphics[width=1.0\textwidth]{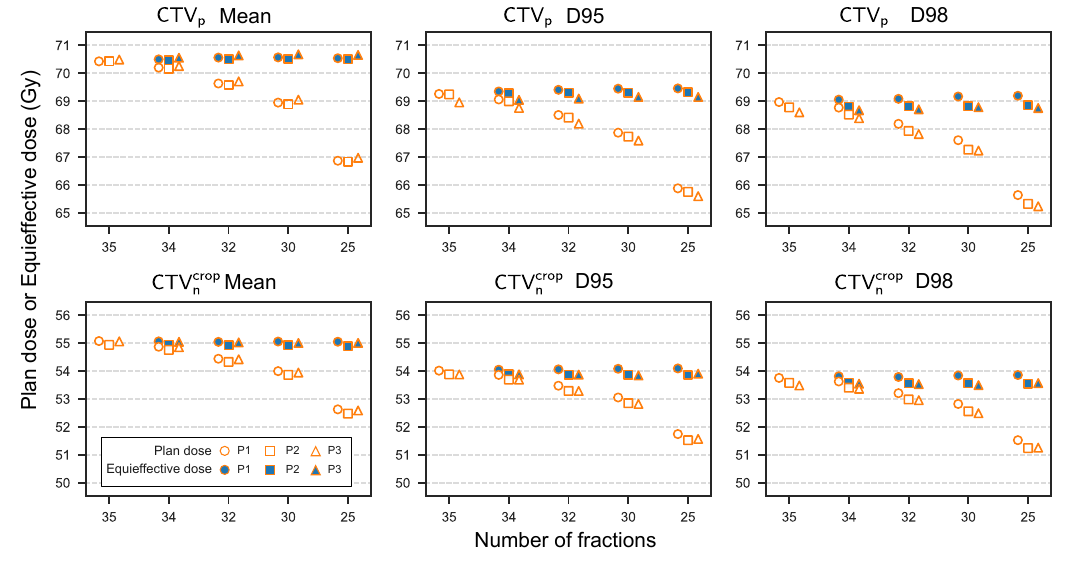}
          \caption{Dose–volume histogram parameters of the planned dose and the calculated equieffective dose $EQD_{n_{ref}}$ for the primary ($\text{CTV}_p$) and secondary clinical target volumes ($\text{CTV}^{crop}_n$, that is $\text{CTV}_n$ cropped 1 cm additionally from the $\text{CTV}_p$). $D_x$ denotes the minimum dose received by $x$\% of the volume.}
          \label{fig:ctv_dose}
        \end{figure}

        Across the patients, an average value of 70.5 Gy was achieved for the mean $\text{CTV}_p$ plan dose in the standard fractionation, which decreased as the number of fractions was reduced in the hypofractionated plans (Figure \ref{fig:ctv_dose}). In contrast, when expressed as $EQD_{n_{ref}}$, the mean $\text{CTV}_p$ dose remained essentially the same for the standard and hypofractionated schedules, with an average value of 70.5 Gy and a standard deviation (SD) of 0.08 Gy. For the secondary target $\text{CTV}^{crop}_n$, the mean dose (and SD) was 55.0 (0.06) Gy. A similar pattern was observed for D95 and D98 in both targets (Figure \ref{fig:ctv_dose}). All standard and hypofractionated treatment plans met clinical goals after conversion to $EQD_{n_{ref}}$.  

\begin{figure}[H]
            \centering
            \includegraphics[width=1.0\textwidth]{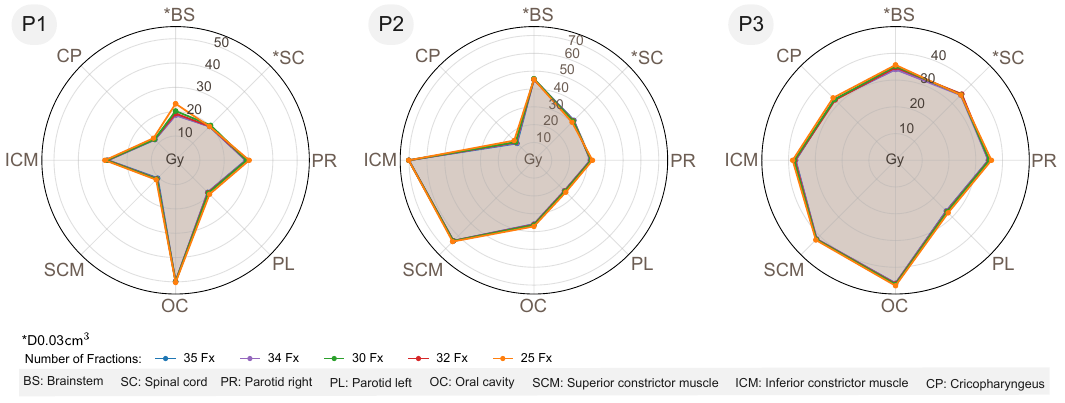}
            \caption{Mean dose and D0.03 $\text{cm}^3$ in terms of equieffective dose $EQD_{n_{ref}}$ (with $n_{ref}$ = 35) for various organs-at-risk in standard and hypofractionation plans.}
            \label{fig:oar_dose_polar}
        \end{figure}

        The OAR doses in the hypofractionated plan were comparable to those in the standard fractionation plan, with the exception of the brainstem maximum dose for patient P1 (Figure \ref{fig:oar_dose_polar}). Nonetheless, all serial organs met the dose constraints for every patient.
    
\begin{figure}[H]
            \centering
            \includegraphics[width=1.0\textwidth]{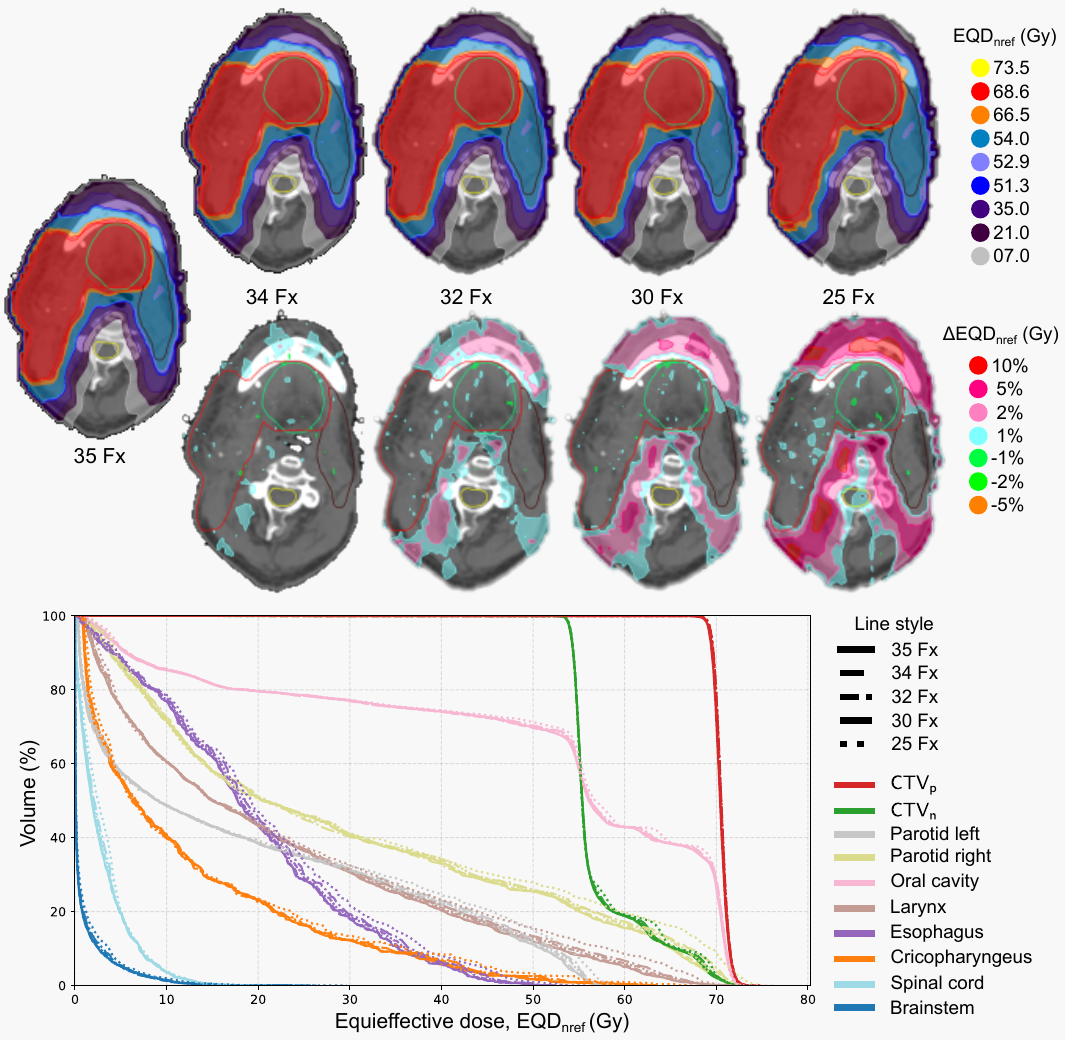}
            \caption{Equieffective dose ($EQD_{n_{ref}}$) distribution, $\Delta EQD_{n_{ref}}$ distribution with respect to standard fractionation and dose-volume histograms for treatment plans in different fractionation regimens. Here, $n_{ref} = 35$ represents the number of fractions in the standard fractionation.}
            \label{fig:patient_dose}
        \end{figure}

        Figure \ref{fig:patient_dose} shows an illustration of the $EQD_{n_{ref}}$ distributions and their corresponding differences for patient P1, together with the DVHs for various ROIs, in both standard fractionation and hypofractionation treatment plans. The target DVHs are nearly identical across the fractionation schemes; however, for the OARs, small but noticeable differences appeared as the number of fractions decreased. DVH plots for the remaining patients are presented in the Supplementary material (Figure \ref{fig:sup_dvh}), together with the DVH parameters for all OARs of the patients (Table \ref{nominalplandvhpara}).

    \subsection{Plan robustness evaluation with equieffective dose}
\begin{figure}[H]
            \centering
            \includegraphics[width=1.0\textwidth]{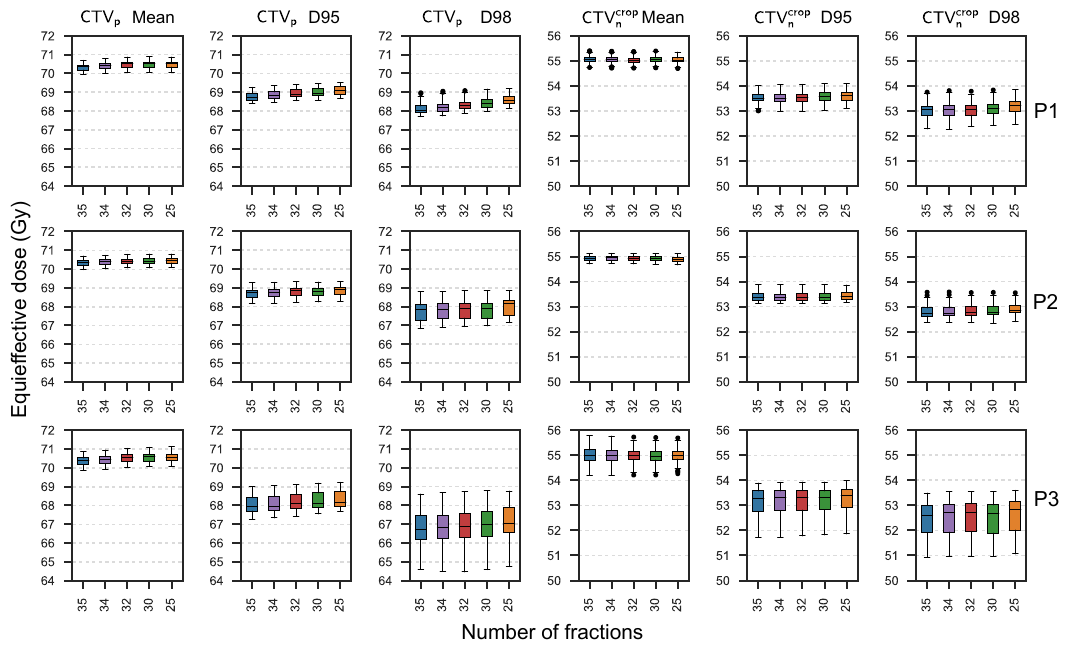}
            \caption{Box plots for various dose-volume histogram parameters expressed as equieffective dose ($EQD_{n_{ref}}$, with $n_{ref}$ = 35) for the primary ($\text{CTV}_p$) and secondary clinical target volume ($\text{CTV}^{crop}_n$, that is $\text{CTV}_n$ cropped 1 cm additionally from the $\text{CTV}_p$) in different robustness scenarios. $D_x$ denotes the minimum dose received by $x$\% of the volume.}
            \label{fig:ctv_robustness}
        \end{figure}

        Compared to standard fractionation, the $\text{CTV}_p$ scenarios in the 34- and 32-fraction plans showed an average $\Delta EQD_{n_{ref}}$ of 0.1 Gy for the DVH parameters across all patients, whereas the 30- and 25-fraction plans resulted in an average difference of 0.2 Gy (Figure \ref{fig:ctv_robustness}). In contrast, the $\text{CTV}^{crop}_n$ scenarios produced an average difference of zero for the DVH parameters. Nevertheless, all scenarios satisfied the robustness criterion of D$95 \geq 95\%$ of the prescribed dose for every patient. By comparison, the stricter requirement of D$98 \geq 95\%$ of the prescribed dose was not achieved for the patient P3. This requirement was satisfied in only 17 scenarios for the 35-, 34-, and 32-fraction regimens, and in 19 and 20 scenarios for the 30- and 25-fraction regimens, respectively. 
        
\begin{figure}[H]
            \centering
            \includegraphics[width=0.4\textwidth]{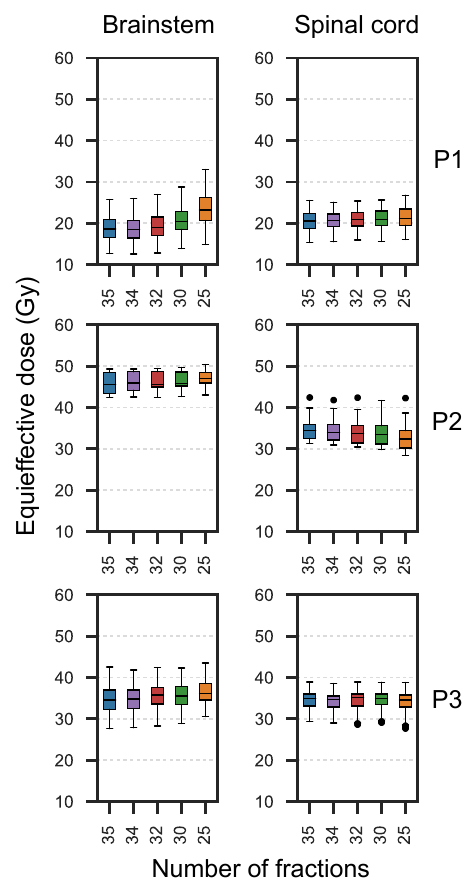}
            \caption{Box plots of D0.03 $\text{cm}^{3}$ expressed as equieffective dose ($EQD_{n_{ref}}$, with $n_{ref}$ = 35) for the spinal cord and brainstem across different robustness scenarios for standard and hypofractionated treatment plans.}
            \label{fig:oar_robustness}
        \end{figure}

\begin{figure}[H]
            \centering
            \includegraphics[width=1.0\textwidth]{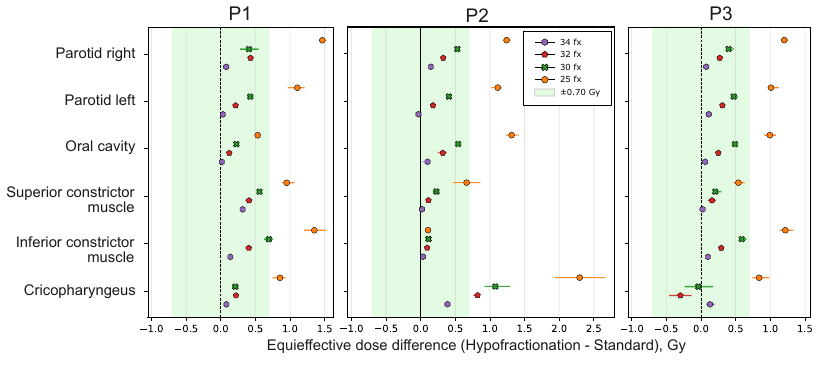}
            \caption{Dose differences ($\Delta EQD_{n_{ref}}$ with respect to standard fractionation) in the mean dose of the parallel organs-at-risk for the hypofractionated plan. The dot indicates the median value, and the line indicates the 90\% bootstrapped confidence interval.}
            \label{fig:forest_oar}
        \end{figure}

        The maximum scenario doses (D0.03 $\text{cm}^3$) to the brainstem and spinal cord were similar between standard fractionation and hypofractionation, except for the brainstem of patient P1 (Figure \ref{fig:oar_robustness}). Nevertheless, for all patients, the D0.03 $\text{cm}^3$ scenario doses to both the brainstem and spinal cord remained well below their respective dose constraints. Furthermore, for the mean dose to the parallel OARs, the median dose difference between the hypofractionated plan of up to 30 fractions and the standard fractionation plan (with a 90\% confidence interval obtained via bootstrapping) was mostly less than 1\% of the prescribed dose (Figure \ref{fig:forest_oar}).  

    \subsection{$\alpha/\beta$ sensitivity analysis}
\begin{figure}[H]
            \centering
            \includegraphics[width=1.0\textwidth]{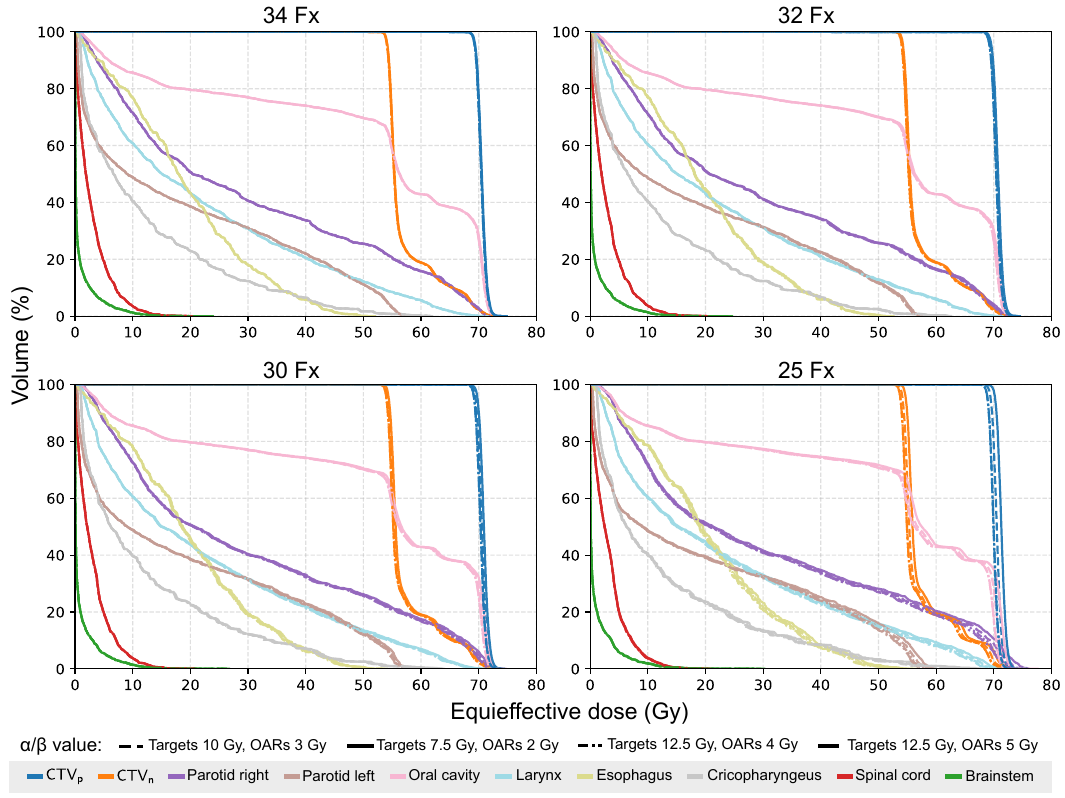}
            \caption{Dose-volume histograms with uncertain $\alpha/\beta$ values for the calculation of equieffective dose ($EQD_{n_{ref}}$, with $n_{ref}$ = 35).}
            \label{fig:dvh_uncertainty}
        \end{figure}

        The effect of uncertainties in the values of $\alpha/\beta$ on the DVH curves of the target and OARs was less evident in the 34-fraction plan, and became somewhat more apparent as the number of fractions were reduced in the hypofractionated plans. As an illustration, the DVH curves for patient P1 are presented in Figure \ref{fig:dvh_uncertainty}. 
        
        Across patients and evaluated target metrics, reducing $\alpha/\beta$ by 25\% resulted in positive $\Delta EQD_{n_{ref}}$ values of 0.04–0.88 Gy, whereas increasing $\alpha/\beta$ by 25\% resulted in negative values of -0.61 to -0.03 Gy (Table \ref{uncertainty_ctv}). The magnitudes were generally larger for the reduced $\alpha/\beta$ value. For targets, the absolute values of $\Delta EQD_{n_{ref}}$ ranged from 0.03 to 0.07 Gy for the 34-fraction regimen and from 0.37 to 0.88 Gy for the 25-fraction regimen.

        For OARs, values lower than reference $\alpha/\beta$ produced positive differences in $EQD_{n_{ref}}$, while higher $\alpha/\beta$ values produced negative differences (Table \ref{uncertainty_oar}). Especially for the brainstem and spinal cord, the difference can go up to 0.63 Gy for a 25-fraction hypofractionated schedule. 
        
        \begin{table}[H]
        \resizebox{\textwidth}{!}{%
        \begin{tabular}{|l|l|cc|cc|cc|cc|}
        \hline
        \textbf{No. of Fractions} &   & \multicolumn{2}{c|}{\textbf{34}} & \multicolumn{2}{c|}{\textbf{32}} & \multicolumn{2}{c|}{\textbf{30}} & \multicolumn{2}{c|}{\textbf{25}} \\
        \hline
        \textbf{Uncertainty $\rightarrow$}&   & -25\%     & +25\%      & -25\%     & +25\%      & -25\%     & +25\%      & -25\%     & +25\%       \\
        \hline
        $\text{CTV}_p$          & Mean dose        & 0.06      & -0.05      & 0.20      & -0.14      & 0.35      & -0.25      & 0.79      & -0.56      \\
        $\text{CTV}_p$          & D98         & 0.06      & -0.05      & 0.20      & -0.13      & 0.34      & -0.23      & 0.77      & -0.54      \\
        $\text{CTV}_p$          & D95         & 0.06      & -0.05      & 0.20      & -0.14      & 0.34      & -0.24      & 0.78      & -0.54      \\
        $\text{CTV}_p$          & D0.03 $\text{cm}^3$       & 0.07      & -0.05      & 0.22      & -0.15      & 0.39      & -0.27      & 0.88      & -0.61      \\
        $\text{CTV}^{crop}_n$   & Mean        & 0.04      & -0.04      & 0.15      & -0.11      & 0.27      & -0.18      & 0.61      & -0.42      \\
        $\text{CTV}^{crop}_n$   & D98         & 0.05      & -0.03      & 0.13      & -0.09      & 0.24      & -0.16      & 0.55      & -0.37      \\
        $\text{CTV}^{crop}_n$   & D95         & 0.05      & -0.03      & 0.14      & -0.09      & 0.24      & -0.16      & 0.55      & -0.38      \\
        \hline
        \end{tabular}%
        }
        \caption{Differences in equieffective dose ($EQD_{n_{ref}}$, with $n_{ref}$ = 35) in Gy for various dose-volume histogram parameters of targets, considering an $\alpha/\beta$ value uncertainty of $\pm25\%$ from the base value of 10 Gy. $\text{CTV}_p$ denotes the primary target, while $\text{CTV}^{crop}_n$ indicates a $\text{CTV}_n$ that is cropped 1 cm additionally from the $\text{CTV}_p$.}
        \label{uncertainty_ctv}
        \end{table}

        \begin{table}[H]
        \resizebox{\textwidth}{!}{%
        \begin{tabular}{|l|l|ccc|ccc|ccc|ccc|}
        \hline
        \textbf{No. of fractions}&   & \multicolumn{3}{c|}{\textbf{34}} & \multicolumn{3}{c|}{\textbf{32}} & \multicolumn{3}{c|}{\textbf{30}} & \multicolumn{3}{c|}{\textbf{25}} \\
        \hline
        \textbf{$\alpha/\beta$ value (Gy) $\rightarrow$}&      &2      &4      &5      &2      &4      &5      &2      &4      &5      &2      &4      &5  \\
        \hline
        Brainstem           & D0.03 $\text{cm}^3$  & 0.04  & -0.03 & -0.05  & 0.13 & -0.08  & -0.15 & 0.24  & -0.16 & -0.27 & 0.55  & -0.37 & -0.63   \\
        Spinal cord          & D0.03 $\text{cm}^3$  & 0.04  & -0.03 & -0.04  & 0.11 & -0.08  & -0.13 & 0.20  & -0.13 & -0.22 & 0.44  & -0.28 & -0.48   \\ 
        Esophagus           & Mean dose & 0.03  & -0.02 & -0.04  & 0.10 & -0.07  & -0.11 & 0.18  & -0.12 & -0.19 & 0.41  & -0.28 & -0.46   \\
        Larynx              & Mean dose & 0.03  & -0.02 & -0.04  & 0.11 & -0.07  & -0.11 & 0.19  & -0.12 & -0.18 & 0.42  & -0.29 & -0.43   \\
        Oral Cavity         & Mean dose & 0.05  & -0.03 & -0.04  & 0.15 & -0.10  & -0.13 & 0.25  & -0.18 & -0.24 & 0.57  & -0.40 & -0.53   \\
        Parotid Left        & Mean dose & 0.03  & -0.02 & -0.03  & 0.09 & -0.06  & -0.10 & 0.16  & -0.11 & -0.18 & 0.36  & -0.25 & -0.41   \\
        Parotid Right       & Mean dose & 0.04  & -0.03 & -0.05  & 0.13 & -0.09  & -0.14 & 0.22  & -0.15 & -0.24 & 0.49  & -0.35 & -0.56   \\
        Cricopharyngeus     & Mean dose & 0.03  & -0.02 & -0.03  & 0.07 & -0.05  & -0.08 & 0.13  & -0.09 & -0.14 & 0.31  & -0.21 & -0.34   \\
        \hline
        \end{tabular}%
        }
        \caption{Differences in equieffective dose ($EQD_{n_{ref}}$, with $n_{ref}$ = 35) in Gy for various dose-volume histogram parameters of organs at risk, considering an uncertain $\alpha/\beta$ values of 2, 4 and 5 Gy with respect to the reference value of 3 Gy.}
        \label{uncertainty_oar}
        \end{table}

        With respect to the prescribed dose, the largest deviation in the $EQD_{n_{ref}}$ parameters due to uncertainty in $\alpha/\beta$ was 1.25\% for the target volumes and 0.90\% for the OARs, with both maxima observed in the 25-fraction plan.

\section{Discussion}
Hypofractionation is a treatment strategy in which the total number of fractions is reduced compared to conventional fractionation, resulting in an increase in the dose per fraction, while maintaining a comparable biological effect. A key methodological choice in this work is to use $EQD_{n_{ref}}$ as the optimization quantity. Unlike $EQD_{X}$ formulations (such as $EQD_{2}$), which normalize the dose to a fixed dose per fraction, $EQD_{n_{ref}}$ preserves a fixed reference number of fractions and thus maintains direct correspondence with conventional clinical prescriptions. Because clinical prescriptions and optimization objectives are typically specified in dose for the reference fractionation, $EQD_{n_{ref}}$ is particularly well suited for exploring and adapting alternative fractionation schedules in relation to the established reference fractionation, as done in the present study. By contrast, $EQD_{2}$-based approaches are generally more appropriate in contexts such as re-irradiation or cumulative dose assessment across multiple treatment courses, where no single reference fractionation scheme exists and a common fraction dose is required \citep{oden2024optimization}.

In this study, we used the $EQD_{n_{ref}}$\textendash based optimization framework to create equieffective hypofractionated IMPT plans for HN patients, reducing the number of fractions by up to 10. However, this strategy is not limited to proton therapy or HN cancer; it can also be applied to photon treatments and other anatomical sites, and in particular to investigate hypofractionation regimens in tumors with low $\alpha/\beta$ values that show higher sensitivity toward fractionation, such as the prostate and breast \citep{aitken2022less, brand2022low}. The biological equivalence between hypofractionated and standard fractionation was determined using $EQD_{n_{ref}}$ dose calculated from the LQ model \citep{bentzen2012bioeffect}.

The hypofractionated plans using $EQD_{n_{ref}}$ met all clinical goals for the targets, and the resulting doses were comparable to those from standard fractionation, with only minor variation in the mean target doses across fractionation regimens. Furthermore, all plans comply with the mandatory clinical goals of the OARs. However, it should be emphasized that the CTV dose in the hypofractionated plan decreased to keep $EQD_{n_{ref}}$ fixed, as the number of fractions were reduced. If the planned dose must remain fixed, or if different dose prescriptions for the targets or different dose constraints for the OARs are required without accounting for the $EQD_{n_{ref}}$, the user can select specific ROIs to be included or excluded from the $EQD_{n_{ref}}$\textendash based evaluation during optimization.

Hypofractionated plans showed smaller dose variations among the robustness scenarios of the primary target. This was driven by the marginally higher coverage and robustness of $\text{CTV}_p$ in the hypofractionated schedules compared to the standard fractionation plan, with a greater number of scenarios meeting the robustness criteria. Nonetheless, given the small difference in the absolute $EQD_{n_{ref}}$ mean dose (Figure \ref{fig:ctv_dose}), the clinical relevance of these findings remains uncertain and should be further investigated. The D0.03 $\text{cm}^3$ doses to the brainstem and spinal cord showed a slight tendency to increase with hypofractionation (Figure \ref{fig:oar_robustness}); however, in all scenarios the values of $EQD_{n_{ref}}$ remained well below the established dose limits. This effect can arise when the dose to an OAR is already well below its defined dose constraint, giving the optimizer more freedom to shape the dose distribution. On the other hand, the optimizer can also reduce the dose, as seen in the spinal cord of patient P2 when the number of fractions decreased (Figure \ref{fig:oar_robustness}). Moreover, the $EQD_{n_{ref}}$ differences in hypofractionated plans arising from uncertain $\alpha/\beta$ values exhibited greater variation with decreasing fractions. Therefore, it is crucial to assess the $EQD_{n_{ref}}$ dose of the plan across different $\alpha/\beta$ values, particularly for serial OARs, to ensure compliance with the OAR dose constraints. 

The rationale for hypofractionation has long been motivated by systemic and patient-centered factors, such as long waiting lists, long travel distances, unexpected treatment interruptions, and the broader goal of reducing cost and improving access to care. At the same time, a persistent shortage of proton therapy resources has underscored the importance of more efficient treatment delivery. Rapid shifts in practice patterns towards increased use of hypofractionation have been reported during the COVID-19 era \citep{thomson2020314, portaluri2020hypofractionation, spencer2021impact}. In this context, the $EQD_{n_{ref}}$\textendash based optimization framework can preserve treatment quality and can be used to evaluate novel biological models or investigate dose-escalation and hypofractionation strategies to enhance treatment efficiency without requiring substantial deviations from established fractionation regimens. This could be particularly beneficial in clinical settings where institutional radiation therapy protocols\textemdash plan design, target dose prescription, and OAR dose constraints\textemdash are originally established for standard fractionation. Since the system automatically accounts for $EQD_{n_{ref}}$ during treatment plan optimization, this avoids the need to manually reformulate dose prescriptions and optimization objectives for each investigated fractionation schedule. Beyond the applications investigated here, the framework could potentially be extended to adaptive radiotherapy workflows, including equieffective re-optimization that accounts for previously delivered dose.

A limitation of the present work is that our $EQD_{n_{ref}}$-based formalism relies purely on the LQ model and does not account for the time component related to the overall treatment time. Overall treatment time is particularly important for fast proliferating tumors \citep{withers1988hazard, fowler1989linear}, such as HN cancer where accelerated proliferation typically triggers after a lag period ($T_k$) of about 3–4 weeks. A carefully regulated hypofractionation schedule that shifts the overall treatment time closer to $T_k$ is advantageous \citep{fowler2001biological}, for example, by reducing the number of fractions from 35 to 25. However, if the time component is omitted from the model, the result would not simply be an equieffective tumor dose, but potentially more effective. In addition, our present analysis of normal-tissue effects focuses primarily on the fractionation sensitivity associated with late-responding tissue component having low $\alpha/\beta$ values, which are largely independent of the time factor; shortening the overall treatment time may negatively affect the acute normal-tissue toxicities, such as mucosal reactions \citep{trott1993time}, that have high $\alpha/\beta$ values.        

The study is also limited by the small number of patients (N = 3), and therefore the results should be interpreted as a proof-of-concept rather than a definitive clinical evaluation.

\section{Conclusion}
Equieffective hypofractionated treatment plans were successfully generated using an optimization framework based on $EQD_{n_{ref}}$. They exhibit minimal differences from conventional fractionated plans in terms of $EQD_{n_{ref}}$ distribution and robustness. The target coverage and the mandatory clinical goals of the OARs were met in all patients. To assess the impact of $\alpha/\beta$ uncertainty on the dose distribution, dose evaluation should consider the full range of plausible $\alpha/\beta$ values for each ROI, particularly when the number of fractions is substantially reduced. In summary, the $EQD_{n_{ref}}$\textendash based optimization framework enabled reductions in the number of treatment fractions while maintaining target coverage, compliance with mandatory OAR clinical goals, and robustness comparable to that of the conventionally fractionated plans.      


\section*{Acknowledgements}
This study was supported by the RAPTOR project which received funding from the European Union’s Horizon $2020$ Marie Skłodowska-Curie Actions under Grant Agreement No. $955956$.

\bibliographystyle{vancouver}
\bibliography{references.bib}

@article{haviland2013uk,
  title={The UK Standardisation of Breast Radiotherapy (START) trials of radiotherapy hypofractionation for treatment of early breast cancer: 10-year follow-up results of two randomised controlled trials},
  author={Haviland, Joanne S and Owen, J Roger and Dewar, John A and Agrawal, Rajiv K and Barrett, Jane and Barrett-Lee, Peter J and Dobbs, H Jane and Hopwood, Penelope and Lawton, Pat A and Magee, Brian J and others},
  journal={The lancet oncology},
  volume={14},
  number={11},
  pages={1086--1094},
  year={2013},
  publisher={Elsevier},
  note={\url{https://doi.org/10.1016/S1470-2045(13)70386-3}}
}

@article{catton2017randomized,
  title={Randomized trial of a hypofractionated radiation regimen for the treatment of localized prostate cancer},
  author={Catton, Charles N and Lukka, Himu and Gu, Chu-Shu and Martin, Jarad M and Supiot, St{\'e}phane and Chung, Peter WM and Bauman, Glenn S and Bahary, Jean-Paul and Ahmed, Shahida and Cheung, Patrick and others},
  journal={Journal of Clinical Oncology},
  volume={35},
  number={17},
  pages={1884--1890},
  year={2017},
  publisher={American Society of Clinical Oncology},
  note={\url{https://doi.org/10.1200/JCO.2016.71.7397}}
}

@article{santos2022role,
  title={The role of hypofractionation in proton therapy},
  author={Santos, Alexandre and Penfold, Scott and Gorayski, Peter and Le, Hien},
  journal={Cancers},
  volume={14},
  number={9},
  pages={2271},
  year={2022},
  publisher={MDPI},
  note={\url{https://doi.org/10.3390/cancers14092271}}
}

@article{bocha2025hypofractionated,
  title={Hypofractionated radiotherapy for head and neck squamous cell carcinoma},
  author={Bocha, Dylan and Le Guevelou, Jennifer and Ortholan, C{\'e}cile and Calderon, Beno{\^\i}t and Pointreau, Yoann},
  journal={Cancer/Radioth{\'e}rapie},
  volume={29},
  number={5-6},
  pages={104681},
  year={2025},
  publisher={Elsevier},
  note={\url{https://doi.org/10.1016/j.canrad.2025.104681}}
}

@article{nilsson2026ultra,
  title={Ultra-hypofractionated versus conventionally fractionated radiotherapy for localised prostate cancer (HYPO-RT-PC): 10-year outcomes of an open-label, randomised, phase 3, non-inferiority trial},
  author={Nilsson, Per and Gunnlaugsson, Adalsteinn and Beckman, Lars and Widmark, Anders and Fransson, Per and Hoyer, Morten and Lagerlund, Magnus and Kindblom, Jon and Johansson, Bengt and Bj{\"o}rnlinger, Kirsten and others},
  journal={The Lancet Oncology},
  volume={27},
  number={3},
  year={2026},
  publisher={Elsevier},
  note={\url{https://doi.org/10.1016/S1470-2045(25)00656-4}}
}

@article{chakraborty2024ultra,
  title={Ultra-Hypofractionated Whole Breast Radiation Therapy Using a Novel Boost Regimen for Early-Stage Breast Cancer},
  author={Chakraborty, MA and Khan, AJ and Tadros, AB and White, C and Zhang, Z and Kim, M and Xu, AJ and LaPlant, Q and O'Brien, DR and others},
  journal={Practical Radiation Oncology},
  volume={16},
  number={1},
  pages={e1--e16},
  year={2026},
  publisher={Elsevier},
  note={\url{https://doi.org/10.1016/j.prro.2025.08.009}}
}

@article{kuncman2025definition,
  title={Definition and requirements for stereotactic radiotherapy: a systematic review},
  author={Kuncman, {\L}ukasz and De la Pinta, Carolina and Milder, Maaike TW and Romero, Alejandra M{\'e}nd{\`e}z and Guckenberger, Matthias and Franzese, Ciro and Scorsetti, Marta and Bruc, Bogumi{\l}a and Verellen, Dirk and Tanadini-Lang, Stephanie},
  journal={Radiotherapy and Oncology},
  volume={211},
  pages={111107},
  year={2025},
  publisher={Elsevier},
  note={\url{https://doi.org/10.1016/j.radonc.2025.111107}}
}

@article{krug2021moderate,
  title={Moderate hypofractionation remains the standard of care for whole-breast radiotherapy in breast cancer: Considerations regarding FAST and FAST-Forward},
  author={Krug, David and Baumann, Ren{\'e} and Combs, Stephanie E and Duma, Marciana Nona and Dunst, J{\"u}rgen and Feyer, Petra and Fietkau, Rainer and Haase, Wulf and Harms, Wolfgang and Hehr, Thomas and others},
  journal={Strahlentherapie und Onkologie},
  volume={197},
  number={4},
  pages={269--280},
  year={2021},
  publisher={Springer},
  note={\url{https://doi.org/10.1007/s00066-020-01744-3}}
}

@article{dearnaley2025moderate,
  title={Moderate hypofractionated radiotherapy for localised prostate cancer: A new standard of care},
  author={Dearnaley, David},
  journal={Cancer/Radioth{\'e}rapie},
  volume={29},
  number={5-6},
  pages={104678},
  year={2025},
  publisher={Elsevier},
  note={\url{https://doi.org/10.1016/j.canrad.2025.104678}}
}

@article{eom2024hypofractionated,
  title={Hypofractionated radiation therapy for head and neck cancers in the era of intensity-modulated radiation therapy},
  author={Eom, Keun-Yong},
  journal={Radiation Oncology Journal},
  volume={42},
  number={1},
  pages={1--3},
  year={2024},
  note={\url{https://doi.org/10.3857/roj.2024.00178}}
}

@article{moon2025hypofractionated,
  title={Hypofractionated radiation therapy in the definitive management of head and neck cancer},
  author={Moon, Dominic H and Awan, Musaddiq J},
  journal = {Seminars in Radiation Oncology},
  volume={35},
  number={2},
  pages={190--196},
  year={2025},
  publisher={Elsevier},
  note={\url{https://doi.org/10.1016/j.semradonc.2025.01.002}}
}

@article{wilkes2006accessing,
  title={Accessing metropolitan cancer care services: practical needs of rural families},
  author={Wilkes, Lesley M and White, Kate and Mohan, Shantala and Beale, Barbara},
  journal={Journal of Psychosocial Oncology},
  volume={24},
  number={2},
  pages={85--101},
  year={2006},
  publisher={Taylor \& Francis},
  note={\url{https://doi.org/10.1300/J077v24n02_06}}
}

@article{emami1991tolerance,
  title={Tolerance of normal tissue to therapeutic irradiation},
  author={Emami, Bahman and Lyman, J and Brown, A and Cola, L and Goitein, M and Munzenrider, JE and Shank, B and Solin, LJ and Wesson, M},
  journal={International Journal of Radiation Oncology* Biology* Physics},
  volume={21},
  number={1},
  pages={109--122},
  year={1991},
  publisher={Elsevier},
  note={\url{https://doi.org/10.1016/0360-3016(91)90171-Y}}
}

@article{marks2010guest,
  title={Guest editor's introduction to QUANTEC: a users guide},
  author={Marks, Lawrence B and Ten Haken, Randall K and Martel, Mary K},
  journal={International journal of radiation oncology, biology, physics},
  volume={76},
  number={3 Suppl},
  pages={S1--S2},
  year={2010},
  publisher={Elsevier},
  note={\url{https://doi.org/10.1016/j.ijrobp.2009.08.075}}
}

@article{bisello2022dose,
  title={Dose--volume constraints fOr oRganS At risk In Radiotherapy (CORSAIR): an “All-in-One” multicenter--multidisciplinary practical summary},
  author={Bisello, Silvia and Cilla, Savino and Benini, Anna and Cardano, Raffaele and Nguyen, Nam P and Deodato, Francesco and Macchia, Gabriella and Buwenge, Milly and Cammelli, Silvia and Wondemagegnehu, Tigeneh and others},
  journal={Current Oncology},
  volume={29},
  number={10},
  pages={7021--7050},
  year={2022},
  publisher={MDPI},
  note={\url{https://doi.org/10.3390/curroncol29100552}}
}

@article{marks2010use,
  title={Use of normal tissue complication probability models in the clinic},
  author={Marks, Lawrence B and Yorke, Ellen D and Jackson, Andrew and Ten Haken, Randall K and Constine, Louis S and Eisbruch, Avraham and Bentzen, S{\o}ren M and Nam, Jiho and Deasy, Joseph O},
  journal={International Journal of Radiation Oncology* Biology* Physics},
  volume={76},
  number={3 Supppl},
  pages={S10--S19},
  year={2010},
  publisher={Elsevier},
  note={\url{https://doi.org/10.1016/j.ijrobp.2009.07.1754}}
}

@article{vreugdenhil2021hypofractionated,
  title={Hypofractionated chemoradiation for head and cancer: Data from the PET NECK trial},
  author={Vreugdenhil, M and Fong, Charles and Sanghera, Paul and Hartley, Andrew and Dunn, Janet and Mehanna, Hisham},
  journal={Oral Oncology},
  volume={113},
  pages={105112},
  year={2021},
  publisher={Elsevier},
  note={\url{https://doi.org/10.1016/j.oraloncology.2020.105112}}
}

@article{fowler1989linear,
  title={The linear-quadratic formula and progress in fractionated radiotherapy},
  author={Fowler, John F},
  journal={The British journal of radiology},
  volume={62},
  number={740},
  pages={679--694},
  year={1989},
  publisher={The British Institute of Radiology},
  note={\url{https://doi.org/10.1259/0007-1285-62-740-679}}
}

@article{paganetti2019report,
  title={Report of the AAPM TG-256 on the relative biological effectiveness of proton beams in radiation therapy},
  author={Paganetti, Harald and Blakely, Eleanor and Carabe-Fernandez, Alejandro and Carlson, David J and Das, Indra J and Dong, Lei and Grosshans, David and Held, Kathryn D and Mohan, Radhe and Moiseenko, Vitali and others},
  journal={Medical physics},
  volume={46},
  number={3},
  pages={e53--e78},
  year={2019},
  publisher={Wiley Online Library},
  note={\url{https://doi.org/10.1002/mp.13390}}
}

@article{oden2025beyond,
  title={Beyond a constant proton relative biological effectiveness: A survey of clinical and research perspectives among proton institutions in Europe and the United States},
  author={{\"O}d{\'e}n, Jakob and Eriksson, Kjell and Kaushik, Suryakant and Traneus, Erik},
  journal={Journal of Applied Clinical Medical Physics},
  volume={26},
  number={1},
  pages={e14535},
  year={2025},
  publisher={Wiley Online Library},
  note={\url{https://doi.org/10.1002/acm2.14535}}
}

@article{mcmahon2019linear,
  title={The linear quadratic model: usage, interpretation and challenges},
  author={McMahon, Stephen Joseph},
  journal={Physics in medicine \& biology},
  volume={64},
  number={1},
  pages={01TR01},
  year={2019},
  publisher={IOP Publishing},
  note={\url{https://doi.org/10.1088/1361-6560/aaf26a}}
}

@article{bentzen2012bioeffect,
  title={Bioeffect modeling and equieffective dose concepts in radiation oncology--terminology, quantities and units},
  author={Bentzen, S{\o}ren M and D{\"o}rr, Wolfgang and Gahbauer, Reinhard and Howell, Roger W and Joiner, Michael C and Jones, Bleddyn and Jones, Dan TL and Van Der Kogel, Albert J and Wambersie, Andr{\'e} and Whitmore, Gordon},
  journal={Radiotherapy and Oncology},
  volume={105},
  number={2},
  pages={266--268},
  year={2012},
  publisher={Elsevier},
  note={\url{https://doi.org/10.1016/j.radonc.2012.10.006}}
}

@article{bolt2021quantification,
  title={Quantification of the uncertainties within the radiotherapy dosimetry chain and their impact on tumour control},
  author={Bolt, Matthew and Clark, Catharine H and Nisbet, Andrew and Chen, Tao},
  journal={Physics and imaging in radiation oncology},
  volume={19},
  pages={33--38},
  year={2021},
  publisher={Elsevier},
  note={\url{https://doi.org/10.1016/j.phro.2021.06.004}}
}

@phdthesis{fredriksson2013robust,
  title={Robust optimization of radiation therapy accounting for geometric uncertainty},
  author={Fredriksson, Albin},
  year={2013},
  school={KTH Royal Institute of Technology},
  address={Stockholm},
  type={PhD dissertation},
  url={https://urn.kb.se/resolve?urn=urn:nbn:se:kth:diva-122262}
}

@article{huang2020hypofractionated,
  title={Hypofractionated radiotherapy alone with 2.4 Gy per fraction for head and neck cancer during the COVID-19 pandemic: The Princess Margaret experience and proposal},
  author={Huang, Shao Hui and O’Sullivan, Brian and Su, Jie and Ringash, Jolie and Bratman, Scott V and Kim, John and Hosni, Ali and Bayley, Andrew and Cho, John and Giuliani, Meredith and others},
  journal={Cancer},
  volume={126},
  number={15},
  pages={3426--3437},
  year={2020},
  publisher={Wiley Online Library},
  note={\url{https://doi.org/10.1002/cncr.32968}}
}

@article{thomson2020314,
  title={Practice recommendations for risk-adapted head and neck cancer radiation therapy during the COVID-19 pandemic: An ASTRO-ESTRO consensus statement},
  author={Thomson, David J and Palma, David and Guckenberger, Matthias and Balermpas, Panagiotis and Beitler, Jonathan J and Blanchard, Pierre and Brizel, David and Budach, Wilfred and Caudell, Jimmy and Corry, June and others},
  journal = {International Journal of Radiation Oncology* Biology* Physics},
  volume = {107},
  pages = {618--627},
  year = {2020},
  note={\url{https://doi.org/10.1016/j.ijrobp.2020.04.016}}
}

@article{portaluri2020hypofractionation,
  title={Hypofractionation in COVID-19 radiotherapy: A mix of evidence based medicine and of opportunities},
  author={Portaluri, M and Barba, MC and Musio, D and Tramacere, F and Pati, F and Bambace, S},
  journal={Radiotherapy and Oncology},
  volume={150},
  pages={191--194},
  year={2020},
  note={\url{https://doi.org/10.1016/j.radonc.2020.06.036}}
}

@article{spencer2021impact,
  title={The impact of the COVID-19 pandemic on radiotherapy services in England, UK: a population-based study},
  author={Spencer, Katie and Jones, Christopher M and Girdler, Rebecca and Roe, Catherine and Sharpe, Michael and Lawton, Sarah and Miller, Louise and Lewis, Philippa and Evans, Mererid and Sebag-Montefiore, David and others},
  journal={The Lancet oncology},
  volume={22},
  number={3},
  pages={309--320},
  year={2021},
  publisher={Elsevier},
  note={\url{https://doi.org/10.1016/s1470-2045(20)30743-9}}
}

@article{bese2007effects,
  title={Effects of prolongation of overall treatment time due to unplanned interruptions during radiotherapy of different tumor sites and practical methods for compensation},
  author={Bese, Nuran Senel and Hendry, Jolyon and Jeremic, Branislav},
  journal={International Journal of Radiation Oncology* Biology* Physics},
  volume={68},
  number={3},
  pages={654--661},
  year={2007},
  publisher={Elsevier},
  note={\url{https://doi.org/10.1016/j.ijrobp.2007.03.010}}
}

@article{grimm2021high,
  title={High dose per fraction, hypofractionated treatment effects in the clinic (HyTEC): an overview},
  author={Grimm, Jimm and Marks, Lawrence B and Jackson, Andrew and Kavanagh, Brian D and Xue, Jinyu and Yorke, Ellen},
  journal={International Journal of Radiation Oncology* Biology* Physics},
  volume={110},
  number={1},
  pages={1--10},
  year={2021},
  publisher={Elsevier},
  note={\url{https://doi.org/10.1016/j.ijrobp.2020.10.039}}
}

@article{niebuhr2021biologically,
  title={Biologically consistent dose accumulation using daily patient imaging},
  author={Niebuhr, Nina I and Splinter, Mona and Bostel, Tilman and Seco, Joao and Hentschke, Clemens M and Floca, Ralf O and H{\"o}rner-Rieber, Juliane and Alber, Markus and Huber, Peter and Nicolay, Nils H and others},
  journal={Radiation oncology},
  volume={16},
  number={1},
  pages={65},
  year={2021},
  publisher={Springer},
  note={\url{https://doi.org/10.1186/s13014-021-01789-3}}
}

@article{oden2024optimization,
  title={Technical note: Optimization functions for re-irradiation treatment planning},
  author={{\"O}d{\'e}n, Jakob and Eriksson, Kjell and Svensson, Stina and Lilley, John and Thompson, Christopher and Pagett, Christopher and Appelt, Ane and Murray, Louise and Bokrantz, Rasmus},
  journal={Medical Physics},
  volume={51},
  number={1},
  pages={476--484},
  year={2024},
  publisher={Wiley Online Library},
  note={\url{https://doi.org/10.1002/mp.16815}}
}

@article{aitken2022less,
  title={When less is more: the rising tide of hypofractionation},
  author={Aitken, K and Mukherjee, S},
  journal={Clinical Oncology},
  volume={34},
  number={5},
  pages={277--279},
  year={2022},
  publisher={Elsevier},
  note={\url{https://doi.org/10.1016/j.clon.2022.03.002}}
}

@article{brand2022low,
  title={How low can you go? The radiobiology of hypofractionation},
  author={Brand, DH and Kirby, AM and Yarnold, JR and Somaiah, N},
  journal={Clinical Oncology},
  volume={34},
  number={5},
  pages={280--287},
  year={2022},
  publisher={Elsevier},
  note={\url{https://doi.org/10.1016/j.clon.2022.02.009}}
}

@article{quashie2023obtaining,
  title={Obtaining organ-specific radiobiological parameters from clinical data for radiation therapy planning of head and neck cancers},
  author={Quashie, Edwin E and Li, X Allen and Prior, Phillip and Awan, Musaddiq and Schultz, Christopher and Tai, An},
  journal={Physics in Medicine \& Biology},
  volume={68},
  number={24},
  pages={245015},
  year={2023},
  publisher={IOP Publishing},
  note={\url{https://doi.org/10.1088/1361-6560/ad07f5}}
}

@book{book2009joiner,
    author = {Joiner, MC and Van  der  Kogel, A},
    title = {Basic  Clinical  Radiobiology},
    publisher = {4th  ed. London: Hodder Arnold},
    year = {2009. p 102-134},
    note={\url{https://phyusdb.wordpress.com/wp-content/uploads/2013/03/basic-clinical-radiobiology.pdf}}
}

@article{stuschke1999fractionation,
  title={Fractionation sensitivities and dose-control relations of head and neck carcinomas: analysis of the randomized hyperfractionation trials},
  author={Stuschke, Martin and Thames, Howard D},
  journal={Radiotherapy and oncology},
  volume={51},
  number={2},
  pages={113--121},
  year={1999},
  publisher={Elsevier},
  note={\url{https://doi.org/10.1016/s0167-8140(99)00042-0}}
}

@article{withers1988hazard,
  title={The hazard of accelerated tumor clonogen repopulation during radiotherapy},
  author={Withers, HR and Taylor, JMG and Maciejewski, B},
  journal={Acta oncologica},
  volume={27},
  number={2},
  pages={131--146},
  year={1988},
  publisher={Taylor \& Francis},
  note={\url{https://doi.org/10.3109/02841868809090333}}
}

@article{fowler2001biological,
  title={Biological factors influencing optimum fractionation in radiation therapy},
  author={Fowler, Jack F},
  journal={Acta oncologica},
  volume={40},
  number={6},
  pages={712--717},
  year={2001},
  publisher={Taylor \& Francis},
  note={\url{https://doi.org/10.1080/02841860152619124}}
}

@article{trott1993time,
  title={The time factor and repopulation in tumors and normal tissues},
  author={Trott, Klaus-R{\"u}diger and Kummermehr, Johann},
  journal={Seminars in Radiation Oncology},
  volume={3},
  number={2},
  pages={115--125},
  year={1993},
  publisher={Elsevier},
  note={\url{https://doi.org/10.1016/S1053-4296(05)80087-6}}
}

\newpage
\appendix

\setcounter{figure}{0}
\setcounter{table}{0}

\renewcommand{\thefigure}{S.\arabic{figure}}
\renewcommand{\thetable}{S.\arabic{table}}

\section{Supplementary material} \label{appendix1}
    
    \begin{figure}[H]
        \centering
        \includegraphics[width=1.0\textwidth]{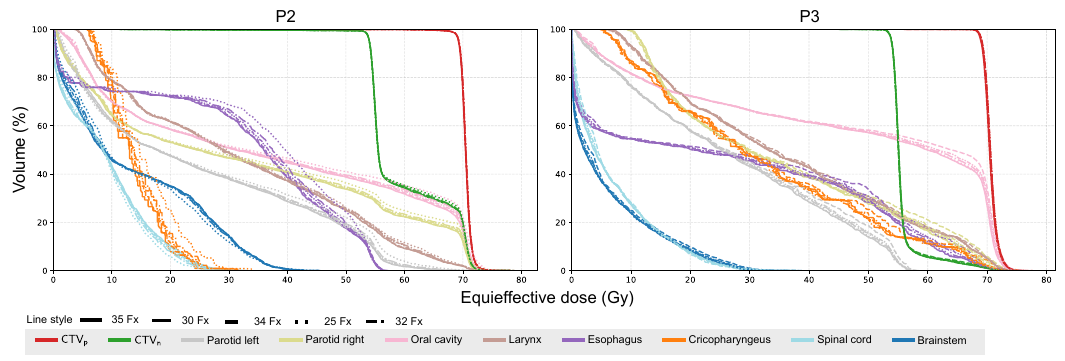}
        \caption{Equieffective dose ($EQD_{n_{ref}}$)-volume histogram for treatment plans with different fractionation schemes in patient P2 and P3. Here, $n_{ref}$ = 35, is the number of fractions in the standard fractionation.}
        \label{fig:sup_dvh}
    \end{figure}

    \begin{table}[H]
        \centering
        \begin{tabular}{|l|l|l|l|l|}
        \hline
        \textbf{OAR Name}          & \textbf{Dose constraints}              & \textbf{P1}   & \textbf{P2}   & \textbf{P3}    \\
        \hline
        Brainstem                  & D0.03 $\text{cm}^3 \leq 54$ Gy (Mandatory*)     & 18.62         & 45.36         & 36.46 \\
        Spinal cord                 & D0.03 $\text{cm}^3 \leq 45$ Gy (Mandatory*)     & 19.69         & 31.73         & 37.89 \\
        Chiasma                    & D0.03 $\text{cm}^3 \leq 54$ Gy (Mandatory*)     & 0.02          & 2.56          & 0.23  \\
        Optic Nerve Left           & D0.03 $\text{cm}^3 \leq 54$ Gy (Mandatory)         & 0.09          & 1.41          & 0.28  \\
        Optic Nerve Right          & D0.03 $\text{cm}^3 \leq 54$ Gy (Mandatory)         & 0.12          & 1.75          & 0.52  \\
        Cochlea Left               & Mean $\leq 45$ Gy (Desirable)          & 0.04          & 46.69         & 0.33  \\
        Cochlea Right              & Mean $\leq 45$ Gy (Desirable)          & 11.00         & 44.36         & 35.21 \\
        Esophagus                  & Mean $\leq 30$ Gy (Desirable)          & 19.04         & 30.65         & 26.20 \\
        Larynx                     & Mean $\leq 40$ Gy (Desirable)          & 21.65         & 31.38         & 36.20 \\
        Oral Cavity                & Mean $\leq 30$ Gy (Desirable)          & 49.77         & 35.79         & 45.98 \\
        Parotid Left               & Mean $\leq 26$ Gy (Desirable)          & 18.65         & 24.20         & 26.79 \\
        Parotid Right              & Mean $\leq 26$ Gy (Desirable)          & 28.82         & 31.63         & 34.67 \\
        Pharyngeal Constrictor Sup & Mean $\leq 55$ Gy (Desirable)          & 10.39         &               & 41.69 \\
        Pharyngeal Constrictor Mid & Mean $\leq 55$ Gy (Desirable)          & 41.49         &               & 60.35 \\
        Pharyngeal Constrictor Inf & Mean $\leq 55$ Gy (Desirable)          & 27.77         &               & 37.33 \\
        Buccal Mucosa Left         & Mean $\leq 30$ Gy (Desirable)          & 0.74          & 21.35         & 20.27 \\
        Buccal Mucosa Right        & Mean $\leq 30$ Gy (Desirable)          & 3.89          & 19.19         & 19.33 \\
        Cricopharyngeus            & Mean $\leq 30$ Gy (Desirable)          & 12.00         & 13.37         & 32.24 \\
        \hline
        \end{tabular}
        \caption{Dose–volume histogram parameters in the patient's nominal plan. ``Mandatory” constraints must be achieved, with constraints marked ``*” must also be prioritized over the target coverage. “Desirable” constraints should be satisfied when possible.}
        \label{nominalplandvhpara}
    \end{table}

\end{document}